\documentclass[smallextended]{svjour3}       
\smartqed  
\usepackage{graphicx}

\usepackage[colorlinks]{hyperref}
\usepackage{gensymb,siunitx} 
\usepackage{booktabs}
\usepackage{cite}
\usepackage{lipsum}
\usepackage{mathtools}
\usepackage{geometry}
\usepackage{ braket,  array}

\begin{document}

\title{Analysis of atmospheric effects on  satellite-based quantum  communication: A comparative study
}


\author{Vishal Sharma         \and
        Subhashish Banerjee 
}


\institute{Vishal Sharma \at
                  IIT Jodhpur, Rajasthan, India.\\ 
                  \email{pg201383506@iitj.ac.in}.          
           \and
          Subhashish Banerjee \at
          IIT Jodhpur, Rajasthan, India.\\
            \email{subhashish@iitj.ac.in}.  
}

\date{Received: date / Accepted: date}

\maketitle

\begin{abstract}
	Quantum Key Distribution (QKD) is a key exchange protocol which is implemented over free space optical links or optical fiber cable. When direct communication is not possible,  QKD is performed over fiber cables, but the imperfections in detectors used at the receiver side and also the material properties of fiber cables limit the long-distance communication. Free-space based QKD is free from such limitations and can pave the way for satellite-based quantum communication to set up a global network for sharing secret messages. To implement free space optical (FSO) links, it is essential to study the effect of atmospheric turbulence. Here, an analysis is made for  satellite-based quantum communication using QKD protocols.
	 We assume two specific attacks,  namely PNS (photon number splitting) and IRUD (intercept-resend with unambiguous discrimination),  which could be main threats for future QKD based satellite applications. The key generation rates and the error rates of the considered QKD protocols are presented. Other parameters such as optimum signal and decoy states mean photon numbers are calculated for each protocol and  distance. Further, in SARG04 QKD protocol with two decoy states, the optimum signal-state mean photon number is independent of the link distance and is valid for the attacks considered here. This is  significant, highlighting its use in a realistic scenario of satellite quantum communication.
\keywords{Free space optics \and  geometric losses \and quantum key distribution \and quantum teleportation \and satellite applications \and space technology \and total attenuation \and turbulence.}

\end{abstract}

\section{Introduction}
\label{intro}
Quantum key distribution \cite{bennett1984quantum,shenoy2017quantum, scarani2009security, srinatha2014quantum} is an advanced secure key exchange technique in the field of quantum communications. Due to high losses, optical fibers are not the practical choice for direct transmission of photons for global distances. Direct satellite links and fiber-based quantum repeaters are the two methods to overcome this problem. Quantum repeater technique will enhance the communication distance significantly which is not possible by optical fibers \cite{sangouard2011quantum, bussieres2013prospective,guha2015rate}. Quantum repeaters based on optical fibers are unable to achieve true global distances and it is also difficult for other approaches based on error correction \cite{munro2012quantum, azuma2015all, muralidharan2014ultrafast}, which need repeater stations placed at intervals of a few kilometers. Therefore, in order to establish communication over global distances many repeater stations are needed, with a large number of qubits per station \cite{boone2015entanglement}.\newline

Quantum secure communication is achieved by three different satellite scenarios. In the first case, a source of entangled photons is implemented on the satellite itself and photons are  sent to two ground stations. This approach helps in distributing two photons to the two users at the same time, separated several thousands of kilometers, even for Lower Earth Orbit (LEO) satellites. After transmission, the correlation property is examined for testing whether the two photons are still entangled or not, in order to confirm the security. Random detection of photons are used for generating the secure key and is  not restricted to the entangled photon security of the source itself. This concept has an important impact on the  satellite based quantum research, where an autonomous satellite  with  an entangled photon source could make  the source functional.
Attenuated laser pulses are the second alternative by which quantum sources can be realized. These laser pulses contain single photons by emitting pulses of low optical power, which results in only a single photon from the source. Decoy pulses must be deployed to avoid the side channel attack due to multi photons per pulse \cite{thapliyal2015applications, pathak2013elements, shukla2014protocols, schmitt2007experimental, lo2005decoy, wang2005beating, ma2005practical, liu2010decoy, pugh2017airborne, liao2017space}. \newline

In the third scenario,  the transmitter and receiver are at the ground, and  satellite station respectively. Hence, here  the signal propagates from Earth to space. This method has a unique feature which includes adapting the quantum source according to the requirement during the complete mission. By this approach,  one can achieve both foundational tests of quantum mechanics and quantum cryptography. In this work, we concentrate on this particular scenario.\newline

The quantum transceiver designed must be small enough to be launched on a  nano-satellite, specially dedicated to this task. A straight forward model would posses  one fixed telescope, around 10 - 30 cm aperture, for sending or receiving photons. A very suitable ground station is needed possessing an optical telescope which tracks the satellites. An optical telescope of a diameter not less than 0.5 m  can be used. In satellite quantum communication, losses are due to diffraction,  which scales more with distance, and not due to absorption. \newline

Satellite-based  quantum communication plays an important and efficient role in the setup of a global network \cite{gisin2002quantum, carbonneau1998opportunities,bennett1992experimental,zbinden2000practical, owens1994photon, hughes2002practical, resch2005distributing, mayers2001unconditional, shields2007key, sharbaf2011quantum, buttler1998practical}. These satellite based quantum communication schemes are designed for FSO communications \cite{kurtsiefer2002long}. For successful implementation of  satellite  based quantum communication, it is necessary to consider free-space QKD under atmospheric turbulence. In an earth-satellite link, only around $30\,\, km$ of the path (depending on the satellite elevation) are inside the atmosphere. The link attenuation must be below 60 \,\,dB for earth to space quantum communication, above this value quantum communication is not feasible. Link distance ($L$) for various scenarios between earth to space are as follows: ground-LEO and LEO-ground links is 500 to 1400 Km; ground-GEO and GEO-ground is above 36, 0000 Km; for LEO-LEO (intersatellite link) is 2, 000 Km; LEO-GEO (intersatellite link) link distance is 35, 500 Km and link distance for GEO-GEO (intersatellite link) is 40, 000 Km.  Although the technological advancement in commercial applications of QKD has met with enormous success,  quantum communication still needs more investigations to deal with issues related to security, data rate,  and communication distance \cite{sharma2016effect, omkar2013dissipative, sharma2014analysis, bedington2017progress,sharma2018decoherence, sharma2018analysis, BazilSharmaBanerjee}. \newline

The Chinese quantum satellite Micius is one of the several Microsatellite  missions launched in the year 2016 which consists of a big platform with a dedicated technology demonstration. This is a space-based quantum key distribution (QKD) system. For the commercial purpose, satellite-based QKD systems must be cost effective, small in size and reliable for real-field applications \cite{khan2018satellite, calderaro2018towards}.The cryptographic key for implementing QKD technology aboard the Chinese satellite Micius,  part of the  quantum experiments at space scale (QUESS) mission placed into orbit in August 2016 and a number of  quantum-optical experiments have been developed and conducted in recent times \cite{khan2018satellite, calderaro2018towards,yin2017satellite, ren2017ground, liao2017satellite, liao2018satellite}.\newline 

There are a number of projects running,  ranging from QKD technology verification within orbit to setting up fully automatic links and key exchange with many ground stations based on optical setup. Some of the relevant examples in this regard are the Japanese SOTA (small optical transponder) laser communication terminal onboard the microsatellite SOCRATES (space optical communications research advanced technology satellite), a hot-air balloon  and photon reflection experiment was performed between  the LAGEOS satellite (laser geodynamics satellite or laser geometric environmental observation survey, using  action of  corner-cube reflectors) and  the Italian Matera Laser-Ranging Observatory (MLRO) as well as a recent Chinese experiment with a small payload on Tiangong-2 Space Lab in the Chinese Micius satellite. Further,  QKD links between ground stations and airplanes have been demonstrated  by many academic groups in  Canada,  Germany, Waterloo and Munich \cite{khan2018satellite, calderaro2018towards, qi2015compact}. \newline 

Currently, most of the projects are aimed towards development of technology. National University of Singapore has investigated entangled-photon on  nanosatellite. QUTEGA is a   German national quantum technologies funding scheme, which will build a nanosatellite to carry a quantum payload with numerous sources embedded in  photonic  chip technique.  The Canadian government is funding  an important project known as  QEYSSat. The aim of the project is to establish a microsatellite into orbit to carry a single photon detection system. Thus,  this is different from other projects, in which a receiver is used in the setup placed in space. This is an important mission which is developed for radiation-hard single-photon detection systems,  polarization-mapping assembly and   a fine-pointing system. Other countries are also performing quantum-based satellite communication projects. Some of them are  CubeSat Quantum Communications Mission started by U.K. and  NanoBob project started by  France and  Austria \cite{khan2018satellite, calderaro2018towards}.\newline 
\newline
The   SpaceQuest experiment, which was jointly developed by the   University of Waterloo and the German aerospace company OHB System, is mainly used for the testing of  quantum-physical effect known as gravitationally induced decoherence. The subsystem was mainly developed by  University of Waterloo, in which quantum key distribution is a secondary mission \cite{khan2018satellite, calderaro2018towards}. In addition to the successful Chinese experiments, several satellite based quantum communication schemes \cite{rarity2002ground, aspelmeyer2003long, nordholt2002present, kurtsiefer2002quantum, hughes2002practical, hughes2002free, pfennigbauer2005satellite, buttler2000daylight, lindenthallong,  fung2008security, toyoshima2008free,toyoshima2008development, toyoshima2009conceptual, resch2005distributing, ursin2007entanglement, villoresi2008experimental, wang2013direct, hughes2000quantum, hughes2004method, bourgoin2013comprehensive, hughes1999secure,  nelson2004system, peloso2009daylight} have also been proposed.\newline 

 In this paper, we have analyzed the performance of Quantum Key Distribution (QKD), in satellite-earth down and up, and in intersatellite links, with two QKD protocols: SARG04 and BB84, with and without decoy states. In real field applications, we have considered real telescope dimensions and usual atmospheric conditions before sunset, 5 dB and 11 dB, in clear summer day. In addition to this, we have considered two specific attacks the photon number splitting,  and the intercept-resend with unambiguous discrimination attacks. We have not included losses due to pointing errors or misalignment of the optics. These effects can be included in the term $\delta_{diff}$, as an additional diffraction (geometric loss). These losses if taken into account,  would only shift, slightly,  the communication distance axis to the right. Various results related to secure key generation rate and communication distance are calculated with and without decoy states for each protocol. In addition to these, effect of mean photon number on secure key generation rate as well as on communication distance is investigated. The  results shown that, it is feasible to establish quantum key distribution with LEO (Low Earth  Orbit) satellites, but not possible with  GEO (Geostationary earth orbit satellites). The  optimum  mean photon number for SARG04 with two decoy states does not depend on distance between transmitter and receiver (link distance), which underscores  its importance to perform well in a real scenario.
\newline
\newline
This paper is organized as follows: Section II sketches the  methodology for an  FSO communication link under various atmospheric conditions. In section III, the secure key rate for different QKD protocols is briefly discussed. We discuss our results in section IV and conclude in section V. 
\section{Methodology for FSO Links under various atmospheric conditions}
	  	    
	  	    It is well known that  three effects mainly  contribute to the total channel attenuation in an FSO link (denoted as $\delta$ $\in$ [0, 1]):diffraction, atmospheric propagation,  and  efficiency of the receiver.

	  	       		    	      Assume that  Cassegrain type telescope architectures  at sender and receiver sides and laser beams of Gaussian type are used for the said arrangement \cite{teich1991fundamentals, alda2003laser}, obscuration and beam diffraction generate attenuation and shown to be  \cite{klein1974optical, degnan1974optical}.	    	    	        
	  	       		    	    	        
	  	       		    	   \begin{eqnarray}
	  	       		    	  	    	    	     \delta_{diff} = (e^{-2\gamma^2_{t}\alpha^2_{t}} -  e^{-2\alpha^{2}_{t}}) (e^{-2\gamma^2_{r}\alpha^2_{r}}  - e^{-2\alpha^{2}_{r}}), \\ \nonumber
	  	       		    	  	    	  \gamma_{t} =\frac{b_{t}}{R_{t}}, \gamma_{r}  =\frac{b_{r}}{R_{r}}, \nonumber 
	  	       		    	  	    	    	   \alpha_{t}  =
	  	       		    	  	    	    	   \frac{R_{t}}{\omega_{t}},\alpha_{r}=\frac{R_{r}}{\omega_{r}},\\\nonumber
	  	       		    	  	    	    	     \omega_{t} =R_{t},\omega_{r} =\frac{\sqrt{2} \lambda L}{\pi R_{t} },\nonumber
	  	       		    	  	        	   \end{eqnarray}  	        
	  	       		    	    	   
	  	       		    	 where $b_{t}$, $b_{r}$,  and $R_{t}$, $R_{r}$ represent radii of the secondary ($b$) and primary ($R$) mirrors at transmitter ($t$) and receiver ($r$) respectively;  L is the distance between telescopes (also known as link distance), $\lambda$ is the considered wavelength  and  $\omega_{t, r}$ is the beam radius at transceiver ends.\newline
	  	       		    	 
	  	       		    	 We are not considering losses due to pointing errors or misalignment of the
	  	       		    	 	   optics. These effects can be included in the term $\delta_{diff}$, as an additional diffraction (geometric losses).  These losses if taken into account,  would only shift, slightly,  the communication distance axis to the right. \newline
	  	       		    	 	    	 	   
	  	       			\begin{figure}[h]
	  	       		    \centering
	  	       		     \includegraphics[width=0.50\textwidth]{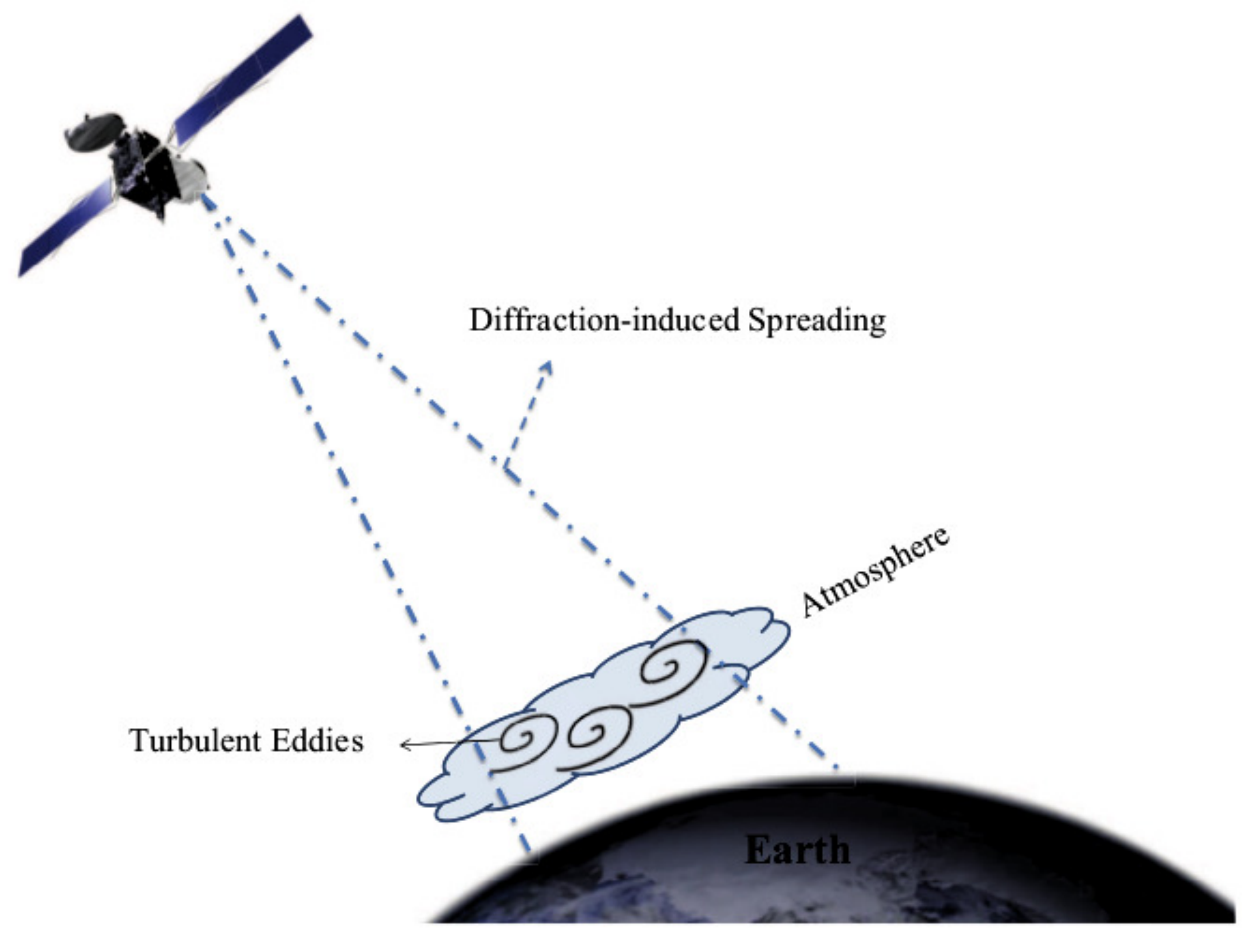}

	  	       				  	       		  	    	    	  	  	    \caption[1]{\textcolor{blue}{Beam-spreading in downlink scenario  \cite{hosseinidehaj2017satellite}}. } \label{downlink} 
 		  	    	    	  	  	    	    	  	 	    	    	  	  	    	    	    \end{figure}

\begin{figure}[h]
	  	       		    \centering
	  	       		     \includegraphics[width=0.50\textwidth]{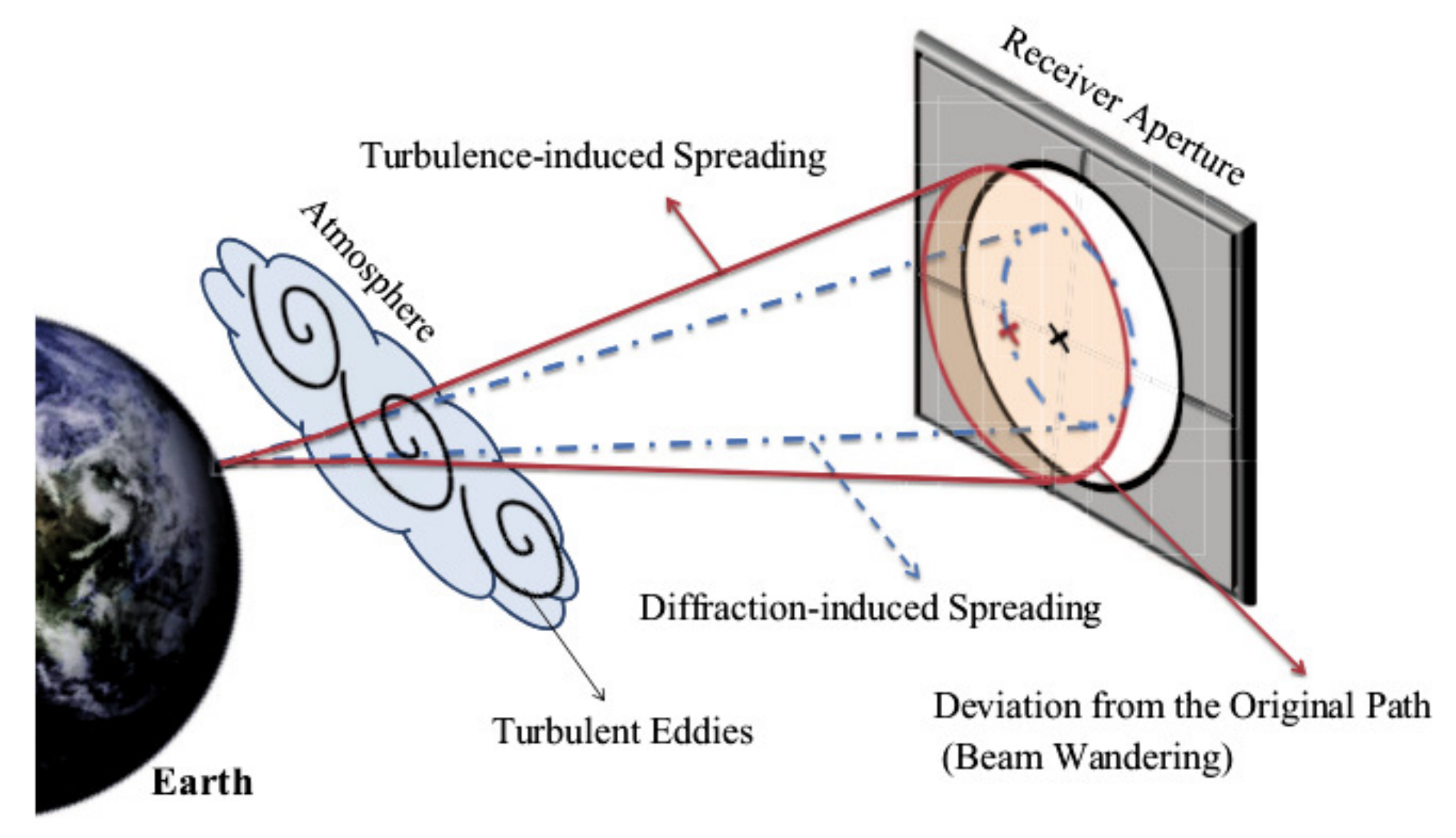}

	  	       				  	       		  	    	    	  	  	    \caption[1]{\textcolor{blue}{Beam-wandering and beam-spreading in uplink scenario  \cite{hosseinidehaj2017satellite}}. } \label{uplink} 
 		  	    	    	  	  	    	    	  	 	    	    	  	  	    	    	    \end{figure}    	 	      		    	 	   	  	       		    	 	   
	  	       		    	 	   
	  	       		    	 	   The atmospheric attenuation  $\delta_{atm}$ is due to various phenomena such as turbulence, scattering and absorption. Hence  it can be written as $\delta_{atm} = \delta_{scatt}\delta_{abs}\delta_{turb}$, where each quantity represents the attenuation of the corresponding phenomena. Here absorption and scattering depend on elevation angle and direction of transmission. The effects due to atmospheric turbulence are enlarged beam divergence, results in less amount of signal power collected by the receive telescope. Other effects generated due to turbulence are decoherence, beam-wander, scintillation and pulse distortion and broadening. The turbulence effects are different for ground to space and space to ground scenarios. In a space to earth scenario light first propagates through vacuum for larger distances before being affected by the atmospheric turbulence, whereas in earth to space scenario, beam spreading effects due to turbulence occur at the beginning of the photon propagation, which causes a high value of divergence. \textcolor{blue}{These generic scenarios are depicted in  Figures \ref{downlink} and \ref{uplink}}.  More detailed description of free space optics and turbulence effects can be obtained from \cite{bloom2003understanding, arnon2003effects,  gabay2006quantum, rarity2002ground, aspelmeyer2003long}. Gas molecules and aerosols absorb the light when it passes through the atmosphere. Turbulence is the main factor which contributes in atmospheric attenuation. This is because of thermal fluctuation which produce refractive index variations. The main factors which determine turbulence are the atmospheric conditions and the position of the ground station \cite{rarity2002ground}. Turbulence effects are calculated by increasing the divergence angle of the beam. In uplink, attenuation caused by turbulence is calculated as \cite{aspelmeyer2003long}
	  	       		    		  	    
	  	       		    		  	     \begin{equation}
	  	       		    		  	     \delta_{turb}= \frac{\Big(\frac{\lambda}{R_{t}}\Big)^{2}}{\Big(\frac{\lambda}{R_{t}}\Big)^{2}+\theta_{turb}^{2}},
	  	       		    		  	     \end{equation}

	  	       		    		  	    where $\theta_{turb}$ is the additional divergence, in radians, produced by turbulence. The expression for  $\theta_{turb}$ is,  $\theta_{turb} = \frac{\lambda}{r_{0}}$, where $r_{0}$ is Fried parameter. ${r_{0}}$ $\approx$ ($\lambda$)$^\frac{6}{5}$. Total channel attenuation is written as
	  	       		    		  	    \begin{equation}
	  	       		    		  	    \delta = \delta_{diff}\delta_{atm}\delta_{rec}.\label{totalattenuation}
	  	       		    		  	    \end{equation}

	  	       		    		  	   The above equation for total attenuation ($\delta$) is represented in dB (dB is calculated as $10 log_{10} (\delta$) = $10 log_{10} (\delta_{diff})$ + $10 log_{10} (\delta_{atm})$ + $10 log_{10} (\delta_{rec}$)\,\,). In above equation $\delta_{diff}$, $\delta_{atm}$ and $\delta_{rec}$ represent attenuation due to geometrical losses, atmospheric losses and losses due to receiver inefficiency, respectively. In our current  work, we are using Eq. \ref{totalattenuation} for calculating total attenuation ($\delta$) which also includes attenuation due to  detector inefficiency. In case of uplink (ground to space links), the total attenuation ($\delta$), excluding attenuation due to detector efficiencies, can also be written as
	  	       		    		  	    \begin{equation}
	  	       		    		  	    \delta = \frac{L^{2}\Big(\theta_{T}^{2}+\theta_{atm}^{2}\Big)}{D_{R}^{2}} \frac{1}{T_{T}\Big(1-L_{P}\Big)T_{R}} 10^{A_{atm}/10},\label{turb}
	  	       		    		  	    \end{equation}
 		    		  	     	    where $A_{atm}$ is the attenuation of the atmosphere in dB. $A_{atm}$ = 1 dB for excellent sight conditions (no haze, fog, or clouds) and is valid only in certain wavelength region. $\theta_{T} = \frac{\lambda}{D_{T}}$, here $\theta_{T}$ is the divergence angle resulting from the transmit telescope. $D_{T}$ is the diameter of the transmit telescope. $L_{P}$ represents pointing loss. $T_{T}$ and $T_{R}$ are the telescope transmission factors. We consider $T_{T}$ = $T_{R}$ = $0.8$. Here we are considering $L_{P}$ = 0. $r_{0}$ is $9\,\, cm$ for $800\,\, nm$. In above equation, $\delta_{rec}$ = $3$ to $3.5\,\, dB$ attenuation must be added which is due to detector efficiency operating in the wavelength range of $650\,\, nm$ to $1550\,\, nm$.  The satellite telescopes radius of the primary and secondary mirrors are $15\,\, cm$ and $1\,\, cm$, respectively. The ground telescope radius of the primary and secondary mirrors are $50\,\, cm$ and $5\,\, cm$, respectively. The values of telescope radii have been obtained from the SILEX Experiment \cite{gatenby1991optical} and the Tenerife's telescope \cite{ursin2007entanglement}. The scattering and absorption attenuation is evaluated using a model of clear standard atmosphere \cite{elterman1964parameters} which results in $\eta_{scatt} = 1\,\, dB$.
 		    		  	\newline

	  	       		    	 For calculating total channel attenuation, the considered parameters are shown in Table \ref{table:nonlinnn}.  We have considered $\lambda =650 nm$, it seems reasonable because suitable avalanche photo detector (silicon avalanche photo detector) for single photon detection is available. At telecom wavelength,  $\lambda =1550\,\, nm$, link attenuation increases due to high beam divergence at large wavelength and due to higher absorption in the atmosphere. At $\lambda =1550\,\, nm$, due to longer wavelength, the photon becomes weaker, hence detection of single photon particularly at this telecom wavelength becomes difficult to detect. The present quantum technology exists between 700-800 nm wavelength range, which is close to visible light and effect of natural light pollution starts dominating. In addition to this, sunlight intensity at 1550 nm is five times weaker than at 800 nm, this is the reason that background noise has to reduce at a very low level, hence it is another difficult task to perform at this telecom wavelength.  	\newline

	  	       		    	    	  	 	   Geometric loss  increases with the increasing link range. In a free space optic model, geometric loss can be reduced by deploying low value of divergence angle of laser beam. Under geometric attenuation,  light beam diverges from transmitter to receiver, hence most of the light beam does not reach  the receiver's telescope and signal loss occurs. It is necessary to increase the receiver aperture area so that geometric losses can be controlled (minimized) by collecting more signal at the receiver telescope.

	    \section{THE SECURE KEY RATE ANALYSIS WITH DIFFERENT
	    PROTOCOLS}
	    
	    Here we study BB84 and SARG04 QKD protocols, with and without decoy states, under two specific attacks, namely PNS (photon number splitting) and IRUD (intercept-resend with unambiguous discrimination) attacks.
	   \subsection{The BB84  protocol}
	   The BB84 protocol was proposed in  \cite{bennett1984quantum}, see \cite{fuchs1997optimal, bruss2000quantum} for details. The attenuated laser pulses used in practical QKD schemes are coherent in nature and described by coherent states. The output pattern obtained from lasers follow the Poisson distribution \cite{teich1991fundamentals, loudon2000quantum}. 
	   	    	       
	   	    	   \begin{equation}
	   	    	   |\alpha\rangle=e^{\frac{-|\alpha|^2}{2}}\sum\limits_{n=0}^\infty\frac{\alpha^n}{\sqrt{n!}}|n\rangle.
	   	    	   \end{equation}
	   	    	   
	   	    	    Here $|\alpha|= \sqrt{\mu},\,\,  \mu$ is the mean photon number of a pulse. The probability $P_{n}(\mu)$ corresponding to $n$ photons in a pulse is given by
	   	    	    
	   	    	   \begin{equation}
	   	    	    P_{n}(\mu)=|\langle n|\alpha\rangle|^2 = e^{-|\alpha|^2}\frac{|\alpha|^{2n}}{n!} = e^{-|\mu|}\frac{|\mu|^{n}}{n!}
	   	    	    \end{equation}	    
	   	    	    
	   	    	     In QKD, the transmitter transmits the bit stream in the form of optical pulses via a quantum channel \cite{ sharma2015controlled,sharma2016comparative}. These optical pulses are specified by a number known as beam intensity $\mu$ (mean photon number) which ranges from 0.1 to 0.5. Here 0.1 indicates 1 photon every 10 pulses \cite{bennett1992experimental, hughes2002practical, resch2005distributing}. For bit encoding in QKD system, the polarization of only a single photon is used. In BB84 protocol, polarization filters are used to polarize the photons \cite{bennett1992experimental,mayers2001unconditional, shields2007key}.	The Shannon mutual information,   I(A : B) and I(B : E),  shared between Alice (A)-Bob (B)  and Bob (B)-Eve (E), respectively are calculated in bits/pulse \cite{scarani2009security, cover2006elements}. Here, 
	   	   		    
	   	    \begin{equation}
	   	    I(A:B) = \sum_{n=0}^{\infty}\Big(1-(1-\delta)^{n}\Big)P_{n}{(\mu)} \approx \mu \delta, \label{IABBB84}
	   	    \end{equation}

	   	    \begin{equation}
	   	    I(B:E) = \sum_{n \geq 2}^{\infty} P_{n}(\mu).\label{IBEBB84}
	   	     \end{equation}
	   	     
	   	     Eve's Information, $I_{Eve}$, is defined as

	   	    \begin{equation}
	   	    I_{Eve} \approx \frac{I(B:E)}{I(A:B)}.\label{IEVE}
	   	    \end{equation}
	   	     
	   	     The lowest value for the key generation rate $R$ (in bits/pulse) is expressed in \cite{fuchs1997optimal,cover2006elements,lo2005decoy}
	   	     
	   	    
	   	    \begin{equation}
	   	    R\geq q \Bigg(-Q_{\mu}f(E_{\mu}) H_{2} E_{\mu}+ \Omega Q_{\mu}\Big(1- H_{2}\Big(\frac{E_{\mu}}{\Omega}\Big) \Big)\Bigg ), \label{RBB84, RSARG04}
	   	    \end{equation}
	   	    where $\Omega$ ($\Omega = 1 - I_{Eve}$)denotes those photons,  from which Eve cannot extract any information, also known as untagged photons \cite{lo2005decoy}. Also $q$ represents the efficiency of the considered protocol, the values of  $q$ are  1/2 and 1/4 for BB84 and SARG04 protocols,  respectively. $f(x)$ represents the bi-directional  error correction efficiency,  whose value is 1.22 for the Cascade protocol \cite{brassard2005quantum}.The yield of the n-photon pulses  is represented as $Y_{n}$ \cite{lo2005decoy}.\newline

	   	     The expected raw key rate can be written as \cite{ma2005practical}
	   	     
	   	    \begin{equation}
	   	    Q_{\mu} = \sum_{n=0}^{\infty} Y_{n} P_{n}(\mu).
	   	    \end{equation}
	   	    
	   	   Quantum Bit Error Ratio (QBER),  $E_{\mu}$, is   \cite{ma2005practical} 
	   	      	    
	   	    \begin{equation}
	   	    E_{\mu} = \frac{\sum_{n =0}^{\infty} Y_{n} P_{n}(\mu)e_{n} }{Q_{\mu}} = \frac{Y_{0}}{2 Q_{\mu}}.
	   	    \end{equation}
	   	       
	    	    Here,  $Y_{0}$ represents dark counts.The above expression is due to the bit error ratio of $n$-photon signals being $e_{n} = \frac{Y_{0}}{2Y_{n}}$,  given that the dark counts, $Y_{0}$, are the only effect causing QBER. Also, for binomial probability distribution,  $\sum_{n =0}^{\infty} P_{n}(\mu) =1$. \newline
	    	         
	    \subsection{The SARG04  Protocol}
	    
	       The SARG04 protocol was proposed in \cite{scarani2004quantum} and is more powerful compared to BB84 against the photon number splitting attack. The quantum communication phase  in SARG04 is similar to that in the BB84 protocol, but the distinction exists in the encryption and decryption of  Shannon's classical information part \cite{scarani2009security}. In this protocol, the bases are not communicated, but Alice declares one nonorthogonal state out of the four pairs  $A_{\omega, \omega^{'}} = \{|\omega x\rangle , |\omega^{'}z\rangle\}$, where $\omega$, $\omega^{'} \in \{+, -\}$ and $|\pm x\rangle = 0 , |\pm z\rangle$ = 1,  \cite{scarani2004quantum, chefles1998unambiguous,  acin2004coherent}.\newline
	       \newline
	        \textbf{\textit{Comparison between SARG04 and BB84 QKD protocols}}\newline
	          
The SARG04 protocol protects against the PNS attack, without extra resources. In the BB84 protocol, Alice and Bob use two non-orthogonal polarization bases Z($|V\rangle$, $|H\rangle$) and X($|45^{\degree}\rangle$, $|-45^{\degree}\rangle$). Alice randomly sends one of the four non-orthogonal polarization states ($|V\rangle$, $|H\rangle$, $|45^{\degree}\rangle$, and $|-45^{\degree}\rangle$). After quantum communication is completed, Alice and Bob each disclose their basis. However, the SARG04 protocol uses four sets: $s_{1}$ = ($|V\rangle$, $|45^{\degree}\rangle$), $s_{2}$ = ($|V\rangle$, $|-45^{\degree}\rangle$), $s_{3}$ = ($|H\rangle$, $|45^{\degree}\rangle$), and $s_{4}$ = ($|H\rangle$, $|-45^{\degree}\rangle$). Alice randomly sends one of the two non-orthogonal polarization states in the randomly selected set; Bob uses either the Z basis or the X basis to measure the photon. After quantum communication is finished, Alice and Bob exchange the information of Alice's selected set and Bob's measuring basis. Thus, quantum communication in SARG04 is identical to the BB84 protocol; only the classical key sifting procedure is modified. At that time, SARG04 is secure for two photon pulses. For example, if Alice sends $|V\rangle$ with two-photon pulses and discloses $s_{1}$ = ($|V\rangle$, $|45^{\degree}\rangle$), then Eve obtains $|V\rangle$ by the Z basis and measures $|45^{\degree}\rangle$ or $|-45^{\degree}\rangle$ with 50 $\%$ probability by the X basis. Eve cannot differentiate the state from her measurement in two-photon pulses. Hence, SARG04 produces a secret key from one-photon and two-photon pulses, whereas BB84 produces a secret key from one-photon pulses. The SARG04 protocol can generate the secret key even when a pulse contains two photons, because Eve cannot get full information of Alice's key from the two-photon pulse \cite{tamaki2006unconditionally}.\newline

	      At first, a photon number non-demolition quantum measurement is performed  during IRUD attack; if a pulse has one or two photons, Eve blocks it, otherwise she performs a generalized quantum measurement. If the measurement is conclusive, she uses a transparent quantum channel to send a copy of the state to Bob. \newline
	       
	       In IRUD attack, Eve introduces some attenuation. If the channel attenuation is less than that introduced by the IRUD attack, Eve should apply a different strategy, otherwise her presence would be immediately detected. In such circumstances, she blocks a fraction $t$ of the single-photon pulses, keeps one photon from each two-photon pulse, and  performs the IRUD attack on the rest of the multi-photon pulses. Then, the attenuation can be written as

	   	    \begin{equation}
	   	   \delta = \frac{(1-t)P_{1}+P_{2}(\mu) + \chi}{\mu},\,\,\,  t \in [0, 1]. \label{Del1SARG04}
	   	    \end{equation}
	   	    	   
	   	    Here $\chi$ is expressed as 
	   	    	   
	   	    \begin{equation}
	   	    \chi = \sum_{n \geq 3}^{\infty} P_{n} (\mu) P_{ok} (n), \label{KaiSARG04}
	   	    \end{equation}
	   	    
	   	    where $P_{ok}$ represents the probability of acceptance. For BB84 protocol, this value is $0.5$ \cite{brassard2005quantum, acin2004coherent}. The value of $P_{ok}$ depends on the number (n) of photons of the state and the overlap of the basis, but it is not less than $\frac{1}{2}$ \cite{chefles1998unambiguous, acin2004coherent, scarani2004quantum}. Using this attack, Eve does not obtain information from single-photon pulses.\newline

	   	   In general, $P_{n}(\mu)$ is the Poisson probability distribution of photons for every weak laser pulse of the transmitter, taking into account the assumption that there are n photons in a pulse. In the same context,  $P_{1}$ and $P_{2}$ represent probability of 1 and 2 photon(s) attenuation.\newline 
	   	    
	   	   In  PNS attack, Eve first performs a photon number non-demolition measurement to identify Alice's multi-photon signals. Eve blocks all single photon pulses, while for multi-photon pulses she stores one photon in a quantum memory, and resends to Bob the remaining photons by a transparent
	   	    quantum channel. When Eve performs the PNS attack, she introduces some
	   	    attenuation. If this attenuation is lower than the channel attenuation, Alice and Bob can not notice the presence of Eve, and thus Eve can obtain full information.  \textit{If the attenuation introduced by PNS attack is more than the allowed channel attenuation, Eve has to apply different strategy to hide her presence.}  \newline

	   	    In SARG04 QKD protocol information is encoded in four nonorthogonal states, when a generalized measure is performed, it is necessary to have at least three copies of the state to obtain a conclusive result with probability $P_{ok}(n)$ \cite{chefles1998unambiguous, buttler1998practical}. Therefore, in order to obtain full information, Eve must carry out an IRUD attack (Intercept-Resend with Unambiguous Discrimination).\newline
	   	    
	   	     \textit{\textcolor{blue}{If the attenuation introduced during }IRUD attack carried out by Eve,  is more than the channel attenuation, her presence can easily be detected by the authenticate users (Alice and Bob). In a  high-attenuation channel, Eve may extract full information about the key. This is the reason that Eve does not apply PNS attack in such conditions. Hence, she applies a different strategy to hide her presence, depicted in Eq. \ref{Del1SARG04}.}\newline

	   	    The attenuation introduced by Eve in Eq. \ref{Del1SARG04} is  defined as the ratio between the mean number of photons that are received and the mean number that would be received in the absence of attenuation (i.e. $\mu$). According to  Eve's planning,  discussed above, the mean number of photons obtained can be calculated as the sum of a fraction (1 - $t$) of the single-photon pulses, plus one photon for each two-photon pulse, plus one photon for the pulses with three or more photons that lead to conclusive measurements. This last term is denoted by $\chi$.\newline

	   	    	 Similarly, Eve blocks a fraction $s$ of the two photon pulses and apply IRUD attack on the remaining multi-photon pulses, to hide herself, as represented by Eq. \ref{Del2SARG04}. The attenuation in this case can be written as	    
	   	   
	   	    \begin{equation}
	   	    \delta = \frac{(1-s) P_{2}(\mu)+\chi}{\mu} ,\,\,\, s \in [0,1]. \label{Del2SARG04}
	   	    \end{equation}
	   	    
	    Fig. \ref{Evesinfo} represents the comparison between $I_{Eve}$ and distance in km under the BB84 and SARG04 protocols. This is calculated based on the link parameters described in subsequent sections. From this figure, it is observed that Eve obtains more information in the BB84 protocol as compared to SARG04 protocol. Hence,  it can be concluded that SARG04 protocol outperforms the BB84 protocol under such specific conditions.\newline
	      
	     In SARG04 protocol, the secret key can be generated with both single photon and two-photon states. The SARG04 protocol can generate the secret key even when a pulse contains two photons, because Eve cannot get full information of Alice's key from the two-photon pulse \cite{tamaki2006unconditionally}, as shown in Fig. \ref{Evesinfo}.\newline

\subsection{Protocols with the decoy-states: An effective approach to counter Eavesdropping}

The decoy-state method was proposed in \cite{hwang2003quantum}, and further studied in \cite{ma2005practical, horikiri2006decoy, wang2005beating}. Introducing decoy-states (also known as extra test states) help in detecting the presence of eavesdropping, whereas signal states are deployed for key generation only \cite{thapliyal2015applications, pathak2013elements, shukla2014protocols}. The shared mutual information is 

\begin{equation}
I(A:B) = P_{1}(\mu) (1-t) + P_{2}(\mu) (1-s) + \sum_{n \geq 3}^{\infty}P_{n}(\mu) P_{ok}(n),
\end{equation}

\begin{equation}
I(B:E) = P_{2}(\mu) (1-s) I_{2} + \sum_{n\geq 3} ^{\infty} P_{n} (\mu) P_{ok}(n),
\end{equation}

here $t$ represents the fraction of the single photon pulses blocked by Eve, and $s$ denotes a fraction of the two-photon pulses.  $I_{2}$ is the amount of information that Eve can obtain from a single copy of the state \cite{scarani2004quantum}. Next, we analyze the security of the protocols  under consideration.\newline

	    \textbf{ 1) BB84 protocol: Vacuum + weak decoy state:}\newline
	   
	    A lower bound on the key generation rate \cite{ma2005practical, meyer2011implement}, based on entanglement distillation described in \cite{gottesman2004security} which in turn use the concept of decoy-state, is 
	    	    
	    \begin{equation}
	   	   R_{BB84} \geq q \Bigg(-Q_{\mu} f(E_{\mu}) H_{2}(E_{\mu}) + Q_{1}\Big(1-H_{2} (e_{1})\Big)\Bigg),\label{RBB84WITHDECOYSTATE}
	   	    \end{equation}
	   	    
	   	    where $Q_{\mu}$  represents the gain of the signal state, $E_{\mu}$ denotes the QBER, $Q_{1}$ represents the gain of single-photon states and $e_{1}$ denotes the error rate of single-photon states.\newline

	    The parameter $Q_{1}$ is  \cite{fung2006performance}
	       
	    \begin{equation}
	    Q_{1} = Y_{1} e^{-\mu} \mu.
	    \end{equation}
	    
	    The lower bound for $Q_{1}$ and upper bound for $e_{1}$ with the vacuum and a weak decoy state ($\nu$) is  \cite{ma2005practical}
	    	        	    
	    \begin{equation}
	    Y_{1}^{L} = \frac{\mu}{(\mu\nu - \nu^{2})} \Bigg(Q_{\nu}e^{\nu} - Q_{\mu}e^{\mu} \Big(\frac{\nu^{2}}{\mu^{2}}\Big) - \frac{(\mu^{2} - \nu^{2})}{\mu^{2}} Y_{0}\Bigg) \leq Y_{1},
	    \end{equation}
	    
	   \begin{equation}
	    Q_{1}^{L} = \mu e^{-\mu} Y_{1}^{L} \leq Q_{1},
	    \end{equation}
	     \begin{equation}
	      e_{1}^{U} =\frac{e_{0}Y_{0}}{ Y_{1}^{L}}\geq e_{1} .
	     \end{equation}

	   \textbf{ 2) The SARG04 protocol: Vacuum + two weak decoy states:}\newline
	   
	   Single-photon states help in key generation rate in BB84 protocol, whereas  both single-photon and two-photon states contribute to the key generation rate in the SARG04 protocol \cite{fung2006performance}. Taking this into account  with the  approach developed in \cite{gottesman2004security}, the gain in case of two-photon pulses is  \cite{cover2006elements,fung2006performance}

 \begin{equation}
  Q_{2} = Y_{2} e^{-\mu}\frac{\mu^{2}}{2}.
  \end{equation}

The SARG04 protocol uses three decoy states,  $\nu_{0}$, $\nu_{1}$ and $\nu_{2}$,  assuming that $\nu_{0}$ is the vacuum (i.e.  $\nu_{0} = 0)$, and the two weak decoy states are  $\nu_{1}$ and $\nu_{2}$.  For these decoy states, gain and quantum bit error rate are \cite{ma2005practical}
\begin{equation}
Q_{\nu_{i}} =\sum_{n=0}^{\infty}Y_{n} P_{n}(\nu_{i}),
\end{equation}

	   \begin{equation}
	   	E_{\nu_{i}} = \sum_{n=0}^{\infty} \frac{Y_{n}P_{n}(\nu_{i})e_{n}}{ Q_{\nu_{i}}}.
	   	  \end{equation}
	   	 The bit error ratio of the n-photon signals, which is  due to only the dark counts  $Y_{0}$, is\newline
	   	 \[e_{n} = \frac{Y_{0}}{2Y_{n}}.\]

	   	 Let the legitimate users (Alice, Bob) select $\nu_{1}$ and $\nu_{2}$ which satisfy \cite{ma2005practical}
	   	 
	   	 \begin{equation}
	   	0 < \nu_{1} < \nu_{2},\,\,\, \nu_{1} + \nu_{2} < \mu.
	   	 \end{equation}

	    Now the key generation can be shown to be \cite{ma2005practical}
	   \begin{equation}
	   \begin{array}{lcl}
	   R_{SARG04} \geq q \Bigg(-Q_{\mu} f(E_{\mu}) H_{2}(E_{\mu}) + Q_{1} \Big(1-H \Big(\frac{Z_{1}}{X_{1}} \Big)\Big)+\\
	    Q_{2} \Big(\Large 1-H(Z_{2}) \Big)\Bigg),\label{RSARG04WITHDECOYSTATE}
	    \end{array}
	   \end{equation}

	 where $X_{n}$ and $Z_{n}$ represents the  binary random variables. $H_{2} (.)$ is the Shannon's binary entropy function \cite{cover2006elements}.\newline

	   \textit{In Eq. \ref{RSARG04WITHDECOYSTATE}, we can replace $H(Z_{1}/X_{1})$ with $H_{2}(e_{1})$, and $H(Z_{2})$ with $H_{2}(e_{2})$, because Eve's attacks (i.e. IRUD attack) does not introduce phase error or bit errors, and we assume that the only source of errors, i.e. the dark counts, is independent of the signal, phase and bit errors which  are independent from each other and have the same distribution.}\newline

	   The  lower limit of the two photon gain is \cite{ma2005practical}

	   \begin{equation}
	  Q^{L}_{2} = \frac{Y^{L}_{2}\mu^2 e^{-\mu}}{2} \leq Q_{2}.
	   \end{equation}

	   The upper limit of $e_{2}$ can be manipulated by considering quantum bit error rate of weak decoy states \cite{ma2005practical}. 
	   	   
	   \begin{equation}
	   E_{\nu_{i}} Q_{\nu_{i}} e^{\nu_{i}} = e_{0}Y_{0} + e_{i}\nu_{i}Y_{1}+e_{2} \frac{\nu^{2}_{i}}{2} Y_{2} + \sum_{n=3}^{\infty}e_{n}Y_{n} \frac{\nu_{i}^{n}}{n!}.
	   \end{equation}

	     \section{ Results}
	    
	    The results shown here are based on three scenarios: uplink, downlink, and intersatellite links. The  parameters for link establishment (shown in Table \ref{table:nonlinnn}) are detector efficiency ($\delta_{rec}$), satellite telescope radius ($R_{t,r}$), satellite secondary mirror radius ($b_{t,r}$), ground telescope radius ($R_{t,r}$), ground secondary mirror radius ($b_{t,r}$), dark counts ($Y_{0}$) \cite{er2005background, bourgoin2013comprehensive} and wavelength ($\lambda$) whose values are 65$\%$, 15 cm, 1 cm, 50 cm, 5 cm, 50 $\times$ $10^{-6}$ counts/pulse and 650 nm,  respectively. $\lambda = 650 \,\,nm$ represents an absorption window with a commercial detector made of silicon avalanche photo diode with high detection efficiency. Silicon avalanche photo diode with internal gain can work with high data rate.  The optical efficiency in the receiver $(f) = \frac{c}{\lambda}$; for 650 nm wavelength, frequency will be 4.61538 $\times$ $10^{8}$ MHz, which is 461.538 THz. The 650 nm region is close to the highest efficiency detection region (65$\%$). The optical frequency (for example of a quasi-monochromatic laser beam) is the oscillation frequency of the corresponding electromagnetic wave. For visible light, optical frequencies are roughly between 400 THz (terahertz =  $10^{12}$ Hz) and 700 THz, corresponding to vacuum wavelengths between 700 nm and 400 nm. We assume a wavelength in the 650 nm region because the diffraction spread is the smallest. Silicon avalanche photo diodes are deployed to detect the wavelengths in between 250 nm and 1100 nm. These photo diodes detect even the very weak light intensities and very fast optical signals because of their avalanche effect. The absorption spectrum of silicon is quite broad. Visible wavelengths (400-1100 $\,\,nm$) are serviced by silicon avalanche photodiode which has $>$ 50 $\%$ detection efficiency with maximum count rates in MHz range and low dark counts. InGaAs avalanche photo diodes and superconducting single photon detectors detect infrared wavelengths (950 - 1650$\,\,nm$). The major drawbacks of InGaAs avalanche photo diodes (APD) are higher dark count rates, lower detection efficiencies and low repetition rates. These are the reasons that InGaAs APDs are not used for satellite mission. 
	    Telescope radius values are taken from \cite{ursin2007entanglement, gatenby1991optical}.  
	
	\begin{figure}[h]
    \centering
     \includegraphics[width=0.50\textwidth]{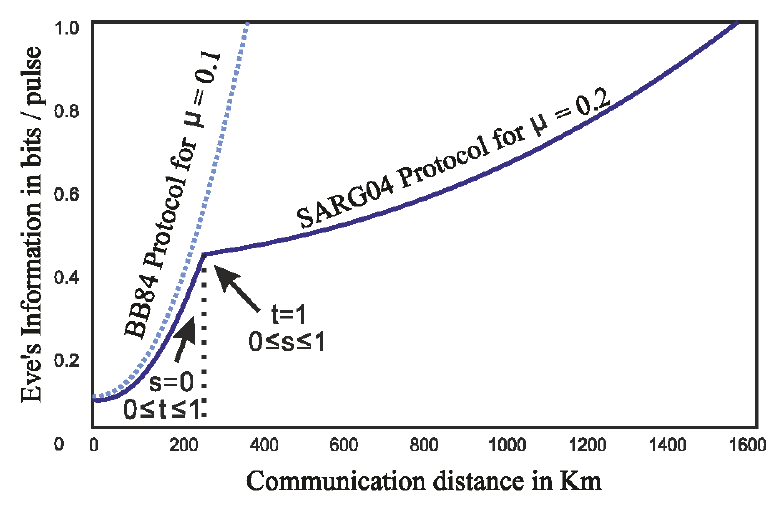}

		  	       		  	    	    	  	  	    	    	  	 	    	    	  	  	    	    	    \caption[1]{Variation in Eve's information with communication distance for each protocol under the uplink case ($\delta_{turb} =5\,\, dB$) calculated using Eqs. \ref{IABBB84}, \ref{IBEBB84}, \ref{IEVE}, \ref{Del1SARG04}, \ref{KaiSARG04}, and \ref{Del2SARG04}. } \label{Evesinfo} 
		  	       		  	    	    	  	  	    	    	  	 	    	    	  	  	    	    	    \end{figure}

The attenuation caused by turbulence in the uplink scenario is computed considering two usual atmospheric scenarios,  one for $\delta_{turb}$ = 5 dB (before sunset) and other for $\delta_{turb}$ = 11 dB (in a clear summer day) \cite{aviv2006laser}. Effect of turbulence on the downlink is almost negligible \cite{rarity2002ground}, \textcolor{blue}{as shown in Tables \ref{table:nonlin}  and \ref{table:nonliin}}. A value of $\delta_{scatt}$ = 1 dB is achieved for the scattering plus absorption attenuation  with the help of Clear Standard Atmosphere model \cite{elterman1964parameters}; these values are similar to those discussed in \cite{rarity2002ground, aspelmeyer2003long}. In Fig. \ref{Evesinfo} we simulated the considered system parameters to interrelate the attenuation with distance and the condition $I_{Eve}$  = 1 is achieved for the optimum parameters (attenuation = 13 dB, $\mu$ = 0.1 for BB84 protocol and attenuation = 25.6 dB, $\mu$ = 0.2 for SARG04 protocol) \cite{scarani2004quantum}. 
	\newline

\begin{figure}[h]	   			  	       		  	    	    	   \centering
    	  \includegraphics[width=0.50\textwidth]{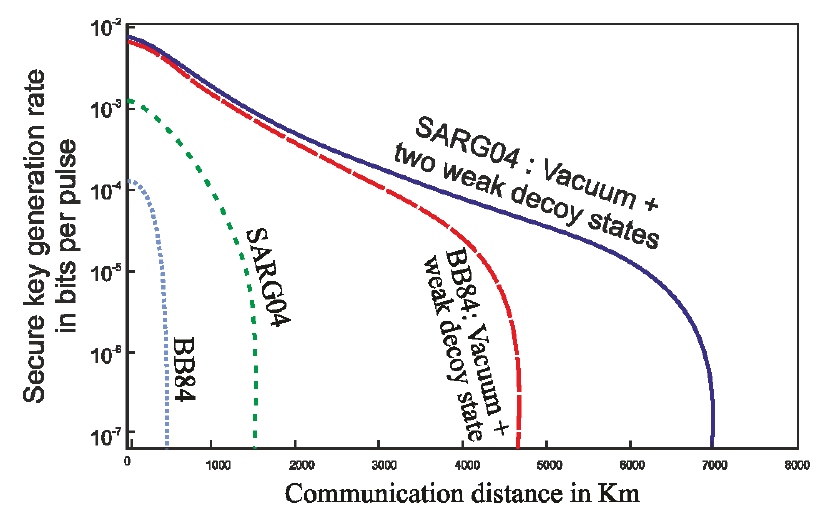}
	   			  	  	    	    	  	 	    	    	  	  	    	    	    	     \caption[1]{In uplink ($\delta_{turb} =5\,\, dB)$, secure key generation rate for all protocols under investigation calculated using Eqs. \ref{RBB84, RSARG04}, \ref{RBB84WITHDECOYSTATE}, and \ref{RSARG04WITHDECOYSTATE}.} \label{securekey}
			  	    	    	    	    	    	    	  	  	    	    	    	    \end{figure}

\begin{figure}[h]
	   	\centering
   	  	 \includegraphics[width=0.50\textwidth]{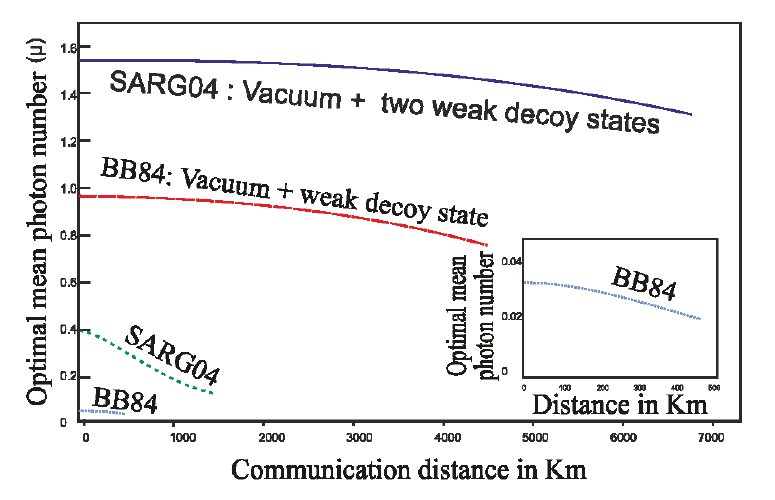}
	   	       	  	       		  	    	    	  	  	    	    	  	 	    	    	  	    	    	  	  	    	    	    	    	    	    
	   	       	  	       		  	    	    	  	  	    	    	  	 	    	    	  	    	    	  	  	    	    \caption[1]{ Variation in optimum mean photon number with communication distance for all protocols under the uplink case ($\delta_{turb} =5\,\, dB)$. These variations in $\mu$ correspond to the highest achievable secure rates, as shown in Fig. \ref{securekey}.} . \label{optmu}
	   	       	  	       		  	    	    	  	  	    	    	  	 	    	    	  	    	    	  	  	    	    \end{figure}

	   	In Fig. \ref{securekey} we have shown the dependency of key generation rates on the communication distance for the considered protocols. The pulses emitted from the laser source can be converted from bits/pulse to bits/second  \cite{schmitt2007experimental}. We take the values of $\mu$ and $\nu$ which are mean photon numbers of signal state and decoy states, respectively, in the range of $[0, 1]$ with a step $0.001$. The number of pulses used as the signal state and the vacuum state are $N_{\mu} = 0.95 N$, and $N_{0}$ = $0.05\,\, N$ (sent by Alice), where $N = 100\,\, Mbit$. In Fig. \ref{optmu}, we have optimized $\mu$ and $\nu^{'s}_{i}$ in each protocol for both the states to obtain the highest key rate.

In Fig. \ref{securekey}, it is observed that critical distance obtained for SARG04 is comparatively higher than BB84, both with and without decoy states.  Also in Fig. \ref{securekey}, it is shown that SARG04 is more robust against eavesdropping than BB84 with an optimal mean photon number. The decoy state method used in BB84 protocol enhances the critical distance. Decoy state method is a powerful technique that increases both the critical distances and key generation rate for both the entangled and non-entangled based protocols \cite{ma2007quantum}.\newline
	   	    	    	    	\newline
	   In case of increasing attenuation, the number of multi-photon pulses must be minimized which helps in reducing the chance of attacks performed by Eve (in this case $\mu$ must be decreased) as shown in Fig. \ref{optmu}. At the higher value of $\mu$, the protocol becomes more robust. With increasing mean photon number, we achieve enhanced communication distance and at the same time, the considered protocols are resistant to Eve's photon number splitting (PNS) attack. Due to movement of the satellite  along its orbit, its distance with the ground station varies. The value of $\mu$ has to be adjusted to achieve the maximum secure rate,  which is the challenging part of the problem. 
	   
	   \begingroup
	    \setlength{\tabcolsep}{2pt} 
	    \renewcommand{\arraystretch}{1.2} 
	    
	     \begin{table*}[ht]
	    	    \caption{  Link Parameters for uplink, downlink and intersatellite links}
	    	     \centering
	    	     
              	    \begin{tabular}{|c| c| }
	            \hline
	      Considered Parameters & Numerical Values \\
	    \hline\hline
	    Detector efficiency ($\delta_{rec}$) & 65$\%$ \\
	    Wavelength ($\lambda$) & 650 \,nm    \\
	    Dark Counts ($Y_{0}$)\cite{er2005background, bourgoin2013comprehensive} & 50 $\times$ $10^{-6}$ counts/pulse \\
	    Ground secondary mirror radius ($b_{t, r}$)    & 5 cm \\
	    Satellite secondary mirror radius ($b_{t, r}$) & 1 cm\\
	    Ground telescope radius ($R_{t, r}$) & 50 cm    \\
	    Satellite telescope radius ($R_{t, r}$) & 15 cm    \\
	    \hline
	      \end{tabular}
	    	     \label{table:nonlinnn}
	    	      \end{table*}
	    	    \endgroup

	  \begingroup
	  \setlength{\tabcolsep}{2pt} 
	  \renewcommand{\arraystretch}{1.2} 
	   \begin{table*}[ht]
	  	    \caption{ Critical distance for different protocols under consideration [Km]}
	  	      \centering
	  	     
	  	          \begin{tabular}{|c| c| c| c| c|}
	  	       \hline
	  	       Scenarios & BB84  & SARG04  & BB84:Vacuum + & SARG04:Vacuum +\\
	  	                &        &         &   weak decoy state  &      two weak decoy state\\	  	              
	  	    \hline
	  	      Downlink & 1540 & 3290 & 9450 & 14100 \\
	  	       Intersatellite  & 430 & 920 & 2660 & 3900 \\
	  	    Uplink($\delta$ = 5 dB) & 460 & 1520 & 4650 & 6980 \\
	  	    Uplink($\delta$= 11 dB)  & - & 500 & 2200 & 3460   \\ [1ex]
	  	      \hline
	  	       \end{tabular}
	  	     \label{table:nonlin}
	  	     \end{table*}
	  	    \endgroup
	  	    
	  	    \begingroup
	  	    \setlength{\tabcolsep}{1pt} 
	  	    \renewcommand{\arraystretch}{1.2} 
	  	  \begin{table*}[ht]
	  	    \caption{ Maximum possible secure rate for different protocols under consideration [Bits/Pulse]}
	  	  	    \centering
	  	    	   
	  	    	    \begin{tabular}{|c| c| c| c| c|}
	  	    	    \hline
	  	    	     Scenarios & BB84  & SARG04  & BB84:Vacuum + & SARG04:Vacuum +\\
	  	    	    	  	                &        &         &   weak decoy state  &      two weak decoy state\\
	  	    	    \hline
	  	    	    Downlink & $1.7 . 10^{-2}$ & $2.4 . 10^{-2}$ & $4.4 . 10^{-2}$ & $4.6 . 10^{-2}$ \\
	  	    	    Intersatellite  & $2.0 . 10^{-2}$ & $2.6 . 10^{-2}$ & $4.8 . 10^{-2}$ & $5.0 . 10^{-2}$ \\
	  	    	    Uplink($\delta$ = 5 dB) & $1.4 . 10^{-4}$ & $1.2 . 10^{-3}$ & $5.8 . 10^{-3}$ & $6.5 . 10^{-3}$\\
	  	    	    Uplink($\delta$= 11 dB)  & - & $7.5 . 10^{-5}$ & $1.4 . 10^{-3}$ & $1.6 . 10^{-3}$   \\ [1ex]
	  	    	    \hline
	  	    	    \end{tabular}
	  	    	    \label{table:nonliin}
	  	    	    \end{table*}
	   				\endgroup  
  	    
\begin{figure}[h]
	   	    	    	    \centering
 	  	    	    	    \includegraphics[width=0.50\textwidth]{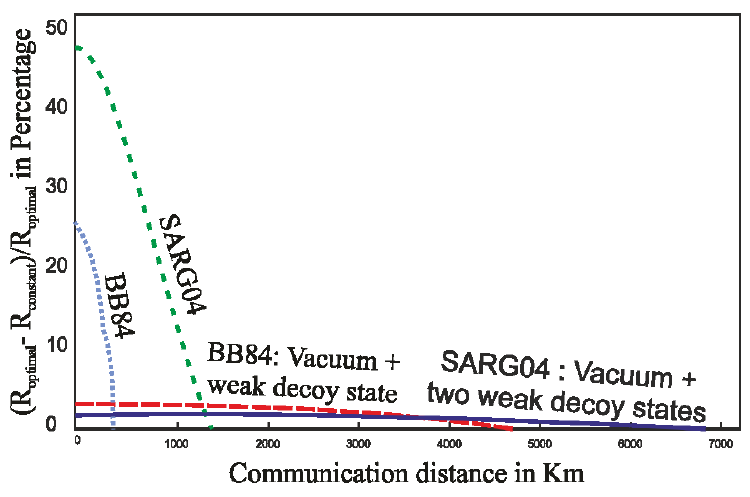}
	   	    	    	    	   	       	  	       	 \caption[1]{(Color online) Blue (dotted) :- BB84; Green (small dashed):- SARG04; Red (large dashed):- BB84: Vacuum+weak decoy state; Purple (continuous dark line):- SARG04: Vacuum+two weak decoy states. In the uplink scenario ($\delta_{turb} =5\,\, dB)$, for each protocol, variation in secure key rate ($ R $) with communication distance at constant value of mean photon number calculated using  Eqs. \ref{RBB84, RSARG04}, \ref{RBB84WITHDECOYSTATE}, and \ref{RSARG04WITHDECOYSTATE}.}\label{optimummu}
	   	    	    	    	   	       	  	       		  	    	    	  	  	    	    	  	 	    	    	  	 	    	    	  	  	    	    	    \end{figure}

	  \textit{In Fig. \ref{optimummu}, for each protocol in uplink scenario, secure key generation rate decreases at constant value of $\mu$, which is independent of the distance}. This is the maximum value at maximum distance for the protocols under analysis. In this figure, for the protocols based on decoy states, we get comparatively low decrease (less than 3$\%$), which clarify that the dependency of $\mu$ on distance is not required. In case of other protocols, keeping $\mu$ constant, secure key rate decreases by $25\%$ and $50\%$ from their maximum values at short distances. The results depicted in Fig. \ref{optimummu},  for each protocol indicates the maximum key generation rates,  keeping  $\mu$ constant to that of optimal $\mu$ for maximum distance. It is clearly observed that in case of protocols based on decoy states the secure rate decreases to a level below  $3\%$, which means that in this situation the variation in mean photon number with distance is not necessary. The result is opposite to that of protocols based on non-decoy states where  rate degradation occurs rapidly.  This implies that the value of mean photon number should vary with distance for obtaining optimum results for secure rates. The rest of the three cases (downlink, uplink in clear weather conditions and inter satellite links) follows the same steps.
	   	    	    	     \newline
	 	  
	   \textit{It would be appropriate, at this juncture, to make a comparison between the results obtained here and the pertinent literature:} \newline

	  1. The total channel attenuation is the result of three effects, it is given by
	  
	  \[\delta = \delta_{diff}\delta_{atm}\delta_{rec}.\].
	  
	  The total channel attenuation is represented by $\delta$. In above equation $\delta_{diff}$ , $\delta_{atm}$ and, $\delta_{rec}$ represent attenuation due to geometrical losses, atmospheric losses and losses due to receiver inefficiency, respectively. We have taken into account this total channel attenuation in our current work, \textit{whereas such type of channel attenuation is not used in} \cite{fung2006performance, tamaki2006unconditionally, jeong2011effects}.\newline
	  
	  2. This work is focused on realistic scenarios of satellite uplink, downlink and intersatellite quantum communication. We are considering two specific type of attacks;  PNS (photon number splitting) attack and \textit{IRUD (intercept-resend with unambiguous discrimination) attacks} \cite{chefles1998unambiguous, buttler1998practical}. The results shown are \textit{valid for these specific attacks}, and hence, our tabulated values differ from that in \cite{ali2010fiber, jeong2011effects},  \textit{where IRUD attack was not considered}.\newline

	  When Eve performs the PNS attack, she introduces some attenuation. If this attenuation is less than the channel attenuation, Alice and Bob can not notice the presence of Eve, and thus Eve can obtain full information. Eve introduces no errors when performing the PNS attack.\newline
	  \newline
	  
	  3. We assume that the dark counts of the detector are the only source of quantum bit errors, which are quantified by the quantum bit error ratio (QBER). In addition to this,  we considered dark count ($Y_{0}) = 50 \times 10^{-6}$ \cite{er2005background, bourgoin2013comprehensive}, which comes in the detection range of the considered detector,  i.e.,  Si-APD (silicon avalanche detector);  whereas value of dark count ($Y_{0}$) is $1.7 \times 10^{-6}$, in \cite{ali2010fiber, jeong2011effects}. We have considered values of the parameters given in Table 1.\newline
	   	 	  
	 4. The lower bound of the key generation rate  for SARG04 is given in Eq. \ref{RSARG04WITHDECOYSTATE} \cite{fung2006performance},  where $X_{n}$ and $Z_{n}$ represent the bit error and the phase error events, respectively, for n-photon pulses. $X_{n}$ and $Z_{n}$ are binary random variables, they take a value equal to one when there is a bit or phase error, and zero othersiwe. H(.) is the Shannon entropy of a random variable \cite{cover2006elements}. \textit{The specific class of Eve's attacks we are considering (i.e.,  IRUD attack) does not introduce phase error or bit errors, and  the only source of errors, i.e.,  the dark counts, is independent of the signal, phase and bit errors which are independent from each other and have the same distribution. Hence, we can replace $H(\frac{Z_{1}}{X_{1}})$ with $H_{2}(e_{1})$, and H($Z_{2}$) with $H_{2}(e_{2})$. Here $e_{1}$ is the error rate of single-photon states.}\newline

	   \textcolor{blue}{Tables \ref{table:nonlin}  and \ref{table:nonliin} exhibit the critical distance and secure key generation rate for both BB84 and SARG04 protocols,  with and without decoy states. The critical distances and secure key generation rates in the downlink are significantly larger as compared to uplink scenario,  due to lack of turbulence induced attenuation in downlink scenario. In addition to this, in MEO satellite downlink quantum communication with decoy states quantum key distribution protocols are possible. These optimum results are not possible in case of intersatellite quantum communication links,   due to reduced telescope dimensions. The most challenging and difficult part in such realistic scenarios are turbulence generated attenuation and telescope dimensions. Comparing all these tabulated results in such realistic scenarios and specific attacks considered here, we conclude that SARG04 QKD protocol with decoy states outperforms the BB84 protocol in terms of both a higher secret key rate and greater secure communication distance.}\newline
	   
	      Geometric attenuation is responsible for the light beam to diverge in its propagation path. To minimize these signal losses,  receiver aperture area is increased to collect more light by the telescope to diminish the geometric losses.  Hence SARG04 protocol deploying with decoy states obtains highest key rate as well as maximum link range. Finally,  we can claim that the optimum results are obtained when we use pulses with two photons plus optimum $\mu$.\newline

In the uplink scenario, secure key generation is low because of high attenuation \cite{bourgoin2013comprehensive, zadok2008secure, bedington2017progress}. At the same time, the value of $\mu$ cannot be increased due to PNS (photon number splitting) attack. To minimize the effects of PNS attack and to increase the secure key generation rate WCP (weak coherent pulse) is preferred over entangled photon source \cite{meyer2011implement, bourgoin2013comprehensive}. The background count rate for uplink is higher than downlink because of artificial light pollution emitted upward \cite{bourgoin2013comprehensive}. Our results show significant design considerations,  e.g.,  type and features of detectors and sources, operating wavelength, ground station locations, specific orbits and telescope design. \newline

Power needed at ground station is more as compared to the power needed at the satellite. The main reason for this is that uplink frequency is set high as compared to downlink frequency. In  uplink, attenuation is more because of turbulence effects. Hence we need powerful devices to send  signals. In addition to this, it is much easier to compensate losses due to attenuation on  earth due to no weight limitation. On the other hand,  weight and space limitations are predominant on the satellite, hence the need to minimize  attenuation \cite{pelton2006basics, manning2009microwave}.\newline

Attenuation and frequency are directly related to each other. Signal losses are higher for higher frequencies, hence more power is required for efficient transmission. The beam of lower frequency is broad whereas a beam of higher frequency is narrow. Earth station has to focus the signal to a small point on satellite in space, which is  performed by deploying a narrow beam generated by higher frequency \cite{rappaport2015wideband, rosen1989satellite}.\newline

Satellite covers a large area on the ground by providing service to many earth stations,  using broad beam generated by lower frequency \cite{gilhousen1990spread, wang2009method}. For visible wavelengths, turbulence effects come into picture when using a transmitter telescope of more than 25-50 cm diameter \cite{pearson1976atmospheric, kedar2004urban}. Turbulence effects can be minimized by selecting a good ground station \cite{rarity2002ground}. In addition to this, an adaptive optics system can be used to minimize the turbulence effects \cite{ricklin2002atmospheric, ellerbroek1994first}.\newline
	    	    
	    \section{Conclusion}
	    
	    We have used realistic scenarios for investigating the feasibility of quantum key distribution, in satellite quantum communication; uplink, downlink and intersatellite links, with BB84 and SARG04 QKD protocols. We have implemented these protocols with and without decoy states for supporting future satellite quantum communication systems. For this reason, we have used practical values of optical hardware and used normal atmospheric conditions. In addition to this, we assumed two specific attacks,  namely PNS (photon number splitting) and IRUD (intercept-resend with unambiguous discrimination),  which could be main threats for future QKD based satellite applications. The key generation rates and the error rates of the considered QKD protocols are presented. Other parameters such as optimum signal and decoy states mean photon numbers are calculated for each protocol and  distance. Our results indicate that, it is possible to establish LEO (Low Earth Orbit) and, MEO (Medium Earth Orbit) satellites. Further, in SARG04 QKD protocol with two decoy states, the optimum signal-state mean photon number is independent of the link distance. This could enhance its significance in a realistic scenario of satellite quantum communication. These results are valid for the \textcolor{blue}{specific} attacks considered here.\newline

	    In order to achieve long distance communication, it is necessary to reduce the link losses. Actual data may be used to better understand  the atmospheric turbulence and define a propagation model  that should help the receiver and transmitter design. Moreover, new communication protocols that exploit the
	    	    	    	    atmospheric turbulence as a resource can be defined. Our telescope design data could be used in future for single photon long distance free space experiments, like teleportation and QKD. This  study will help to experimentally demonstrate the feasibility of Earth-space quantum links.\newline 
	      
	    	      The uplink allows the complex quantum source to be kept on the ground while only simple receivers are in space, but suffers from high link loss due to atmospheric turbulence, necessitating the use of specific photon detectors and highly tailored photon pulses.  For better performance and to enhance the communication distance one could use six or more nonorthogonal states. Further,  the effect of adding pointing and misalignment errors need to be taken into account for greater improvement.\newline
	    	      
	    	      Downlink performance is better than uplink scenario, the reason being that we can place heavy receiving telescopes on earth as compared to space. Also most of the time the beam propagates in vacuum with small diffraction spreading and comes under the effect of atmospheric turbulence in the final stage of propagation.\newline
	    	      
	    	      In this work low earth orbits (altitude upto 1000 km) are considered. They provide advantages of lower optical loss, less costly to attain and easy to operate than higher orbits, making them feasible in near future. To reduce background noise,  quantum key distribution link can be performed at nighttime. Hence,  one can aim to achieve a global scale quantum key distribution.

 \section*{Acknowledgements} 
		
		VS would like to thank the Ministery of Human Resource Development, Govt. of India, for offering a doctoral fellowship as a Ph.D. research scholar at Indian Institute of Technology Jodhpur, Rajasthan, India. VS thanks,  Professor K. K. Sharma for useful discussions pertaining to the work.

\end{document}